\documentclass[12pt,reqno]{amsart}
\usepackage{amsmath,amsfonts,amsthm,amssymb}
\newcommand\version{June 3, 2020}
\usepackage{amsxtra,graphicx,hyperref}

\newtheorem{theorem}{Theorem}
\newtheorem{proposition}{Proposition}
\newtheorem{lemma}{Lemma}
\newtheorem{corollary}{Corollary}

\theoremstyle{definition}

\theoremstyle{remark}
\newtheorem{remark}{Remark}

\numberwithin{equation}{section}

\newcommand{\B}{\mathfrak{B}}
\newcommand{\C}{\mathbb{C}}

\newcommand{\D}{\mathfrak{D}}

\renewcommand{\epsilon}{\varepsilon}

\newcommand{\N}{\mathbb{N}}

\renewcommand{\phi}{\varphi}
\newcommand{\R}{\mathbb{R}}

\newcommand{\Hh}{\mathcal{H}}
\newcommand{\hf}{\mathfrak{h}}

\DeclareMathOperator{\infspec}{inf\, spec\,}

\newcommand{\rmi}{\mathrm i}
\newcommand{\rme}{\mathrm e}

\newcommand{\beq}{\begin{equation}}
\newcommand{\eeq}{\end{equation}}
\newcommand{\half}{\mbox{$\frac{1}{2}$}}

\begin{document}

\title[Emergence of Haldane pseudo-potentials  --- \version]{Emergence of Haldane pseudo-potentials\\ in systems with short-range interactions}

\author{Robert Seiringer}
\address{Robert Seiringer, IST Austria, Am Campus 1, 3400 Klosterneuburg, Austria}
\email{robert.seiringer@ist.ac.at}

\author{Jakob Yngvason}
\address{Jakob Yngvason, Faculty of Physics, University of Vienna, Boltzmanngasse 5, 1090 Vienna, Austria}
\email{jakob.yngvason@univie.ac.at}

\thanks{\copyright\, 2020 by the authors. This paper may be  
reproduced, in
its entirety, for non-commercial purposes.}

\begin{abstract} 
In the setting of the fractional quantum Hall effect we study the effects of strong, repulsive two-body interaction potentials  of short range. We prove that Haldane's pseudo-potential operators, including their pre-factors, emerge as mathematically rigorous limits of such interactions when the range of the potential tends to zero while its strength tends to infinity. 

In a common approach the interaction potential is expanded in angular momentum eigenstates in the lowest Landau level, which amounts to taking the pre-factors to be the moments 
of the  potential. Such a procedure is not appropriate for very strong interactions, however, in particular not in the case of  hard spheres. We derive the  formulas valid in the short-range case, which involve the scattering lengths of the interaction potential in different angular momentum channels rather than its moments. Our results hold for bosons and fermions alike  and generalize previous results in \cite{LS}, which apply to bosons in the lowest angular momentum channel.

Our main theorem asserts the convergence in a norm-resolvent sense of the Hamiltonian on the whole Hilbert space, after  appropriate energy scalings, to  Hamiltonians with contact interactions in the lowest Landau level. 
\end{abstract}

\date{\version}

\maketitle

\section{Introduction}

In a seminal paper \cite{H} on the fractional quantum Hall effect\footnote{A standard reference on the quantum Hall effect is \cite{PG}; see also the reviews \cite{BF} and  \cite{Tong}.}   F.D.M.\ Haldane introduced two-body interaction operators (pseudo-potentials) that have Laughlin's wave functions \cite{Lau} as exact eigenstates. In suitable units the latter are the functions
\beq\label{def:Lau}
\Psi^{\rm L}_{m} (x_1,\dots,x_N) = C_{N,m} \prod_{i<j} (z_i-z_j)^{m} \prod_i \rme^{-|x_i|^2/2}
\eeq
where $N\geq 2$ is the particle number, the $z_j=x^{(1)}_j+\mathrm i x^{(2)}_j\in \mathbb C$ with $x_j=(x^{(1)}_j, x^{(2)}_j)\in\mathbb R^2$ are complex coordinates for the particles moving in a two-dimensional plane perpendicular to a strong magnetic field, $m$ is a positive integer (even for bosons, odd for fermions), and $C_{N,m}$ a normalization factor. The bosonic case, considered in \cite{LS} in the lowest angular momentum channel,  is in particular relevant for dilute quantum gases in rapid rotation, where the rotational velocity takes the role of the magnetic field \cite{C,RSY}. 

Haldane's pseudo-potential operators have the form
\beq\label{def:Pell} \sum_{i\neq j}\mathfrak P^{(\ell)}_{ij}\eeq
where $\mathfrak P^{(\ell)}_{ij}$, for a nonnegative integer $\ell$,  is the projector onto states in the lowest Landau  level (LLL)  with relative angular momentum $\ell$ for a pair $ij$.
Recall that $N$-particle wave functions in the LLL are functions in $L^2(\mathbb R^{2N})$ of the form
\beq \Psi (x_1,\dots,x_N)=\psi(z_1,\dots, z_N)\prod_i \rme^{-|x_i|^2/2}\eeq
with analytic functions $\psi$. They are the lowest energy eigenfunctions of the magnetic kinetic energy part $H^{(0)}$ of the Hamiltonian \eqref{hamiltonian} introduced below. Such a function has relative angular momentum $\ell$ with respect to a pair $ij$ if
\beq \psi(z_1,\dots, z_N)=(z_i-z_j)^\ell \varphi(z_1,\dots, z_N)\eeq
with  $\varphi$ depending  on $z_i$ and $z_j$ only in the combination $z_i+z_j$.
A general analytic function $\psi$ can be expanded in powers of the difference variable $z_i-z_j$ with $z_i+z_j$ and $z_k$, $k\neq i,j$, fixed, and 
$\mathfrak P^{(\ell)}_{i,j}$ picks out the $\ell$th derivative w.r.t.\ $(z_i-z_j)$ at zero, annihilating the other terms. Hence
$\mathfrak P^{(\ell)}_{i,j}$ can be regarded as a zero-range interaction. In fact, as a quadratic form on states in the LLL with relative angular momentum $\geq \ell$,   $\mathfrak P^{(\ell)}_{i,j}$ is formally equal to $2\pi (2^\ell \ell!)^{-1} \Delta^{\ell} \delta(x_i-x_j)$ with $\Delta=\nabla^2$ the Laplacian  
\cite{Pap,Maschk}. 
A Laughlin wave function $\Psi^{\rm L}_{m}$ is a zero energy ground state of \eqref{def:Pell} for all $m\geq \ell+1$.

 The single-particle Hilbert space $L^{2}(\mathbb R^2)$ splits into Landau levels corresponding to the eigenvalues $4n$, $n=0,1,\dots$ of the single particle magnetic kinetic energy \eqref{def:h}. 
The full $N$-particle Hilbert space $\mathcal H=L^2(\mathbb R^{2N})$ (or its antisymmetric or symmetric part, for  fermions or bosons respectively) is a direct sum of the space where all particles are in the lowest Landau level, denoted LLL as above, and its orthogonal complement, where some particles are in higher Landau levels.
 
 Contrary to \eqref{def:Pell}, {\it bona-fide} potentials, defined as multiplication operators by measurable functions of $x\in\mathbb R^2$, do not leave the LLL invariant but generate also states in higher Landau levels.
In 
 \cite{H} the pseudo-potentials were obtained by expanding the projection 
 onto the lowest Landau level of a radial  interaction potential $v(x_i-x_j)$   into angular momentum eigenstates, i.e, writing
 \beq\label{Hpp}P_{\rm LLL} v(x_i-x_j)P_{\rm LLL}=\sum_{\ell\geq 0}\langle \varphi_\ell |v|\varphi_\ell\rangle \mathfrak P^{(\ell)}_{i,j}\eeq
 where $P_{\rm LLL}$ is the projector onto the LLL 
 (in all variables) and
 \beq\label{def:phiell}  \varphi_\ell (z)=(\pi\ell !)^{-1/2} z^\ell \rme^{-|z|^2/2}\eeq
 the single-particle angular momentum eigenfunctions in the lowest Landau 
 level.\footnote{The term \lq\lq pseudo-potentials\rq\rq\ is often used for the expansion coefficients in \eqref{Hpp} rather than the projection operators \eqref{def:Pell}.} In the simplest case, $\ell=0$, the expansion coefficient is 
essentially the integral $\int v$ (except for a Gaussian factor), and for higher $\ell$ it is proportional to the $2\ell$th moment, $\int r^{2\ell}\,v $. It is, however,  clear that this is not viable
for very strong interaction potentials, in particular not if $v$ has a genuine hard core so that all its moments are infinite. In order for \eqref{Hpp} to be meaningful it is necessary that the quadratic form domain of the multiplication operator $v(x_i-x_j)$ has a nontrivial intersection with the LLL.

To obtain valid formulas in the limit when the range of the interaction tends to zero but its strength to infinity, it is necessary to take the kinetic energy operator in higher Landau levels into account. Indeed, the local structure of wave functions in the LLL is quite restricted due to analyticity. Wave functions  with finer structure, avoiding configurations where a {\em bona fide} interaction potential is very large, must necessarily have components also in higher Landau levels. Although these components may tend to zero as the range tends to zero, the joint effects of the kinetic and potential energies at length scales much smaller than the magnetic length\footnote{The magnetic length for a charge $e$ in a field of strength $B$  is $\sqrt{\hbar/eB}$. In our units  $\hbar=e=1$ and $B=2$, so the magnetic length is $1/\sqrt 2$.} will leave a trace in the LLL, and lead in particular to a replacement of $\int v$ as the coupling constant in front of \eqref{def:Pell} by a constant proportional to the $s$-wave scattering length of $v$, as rigorously established in \cite{LS}. 

In the present paper we generalize the analysis in  \cite{LS} to include all angular momentum channels. This allows in particular also to treat fermions, where only odd angular momenta occur.
To account for the fact that strong, short range interactions may create states with arbitrarily large energy in ever higher Landau levels it is convenient to consider convergence of operators in resolvent sense. An elegant way to achieve this uses the concept of {\em $\Gamma$-convergence} \cite{masi} which involves resolvents of operators. The resolvents of Hamiltonians with strong interaction suppress states with very high energy and eventually, in the limit considered, only states in the LLL survive. 

Our main theorem states that after suitable $\ell$-dependent energy scaling the full Hamiltonian converges in this sense  to a Hamiltonian in the lowest Landau level with the pseudo-potential interaction operator \eqref{def:Pell} and a definite coupling constant. Generalizing the $\ell=0$ case, the coupling constant is  determined by the $\ell$-wave scattering length of $v$, which can be obtained from a variational principle. It is essential here to note that the effective Hamiltonian in the LLL is {\em not} obtained by projecting the original Hamiltonian onto the LLL. The effective coupling constants are renormalized by properly taking into account the behavior of the system at length scales much shorter than the magnetic length. In the  limit of zero range/infinite strength one obtains a Hamiltonian that operates within the LLL with the renormalized coupling constants.

In contrast to \cite{LS} we shall mainly focus on  strictly two-dimensional systems for simplicity,  but in Section~\ref{sec:3d} we discuss the modifications that are necessary to apply our results to three-dimensional systems with a  confining potential in the third direction. To keep  the proofs simple and  transparent, 
we do not keep track of the $N$-dependence of our estimates but rather regard the particle number as fixed. Quantitative, $N$-dependent estimates as derived in \cite{LS} for $\ell=0$ would also be possible with some additional work. The estimates in \cite{LS} are far from optimal, however,  and we leave it as an open problem to improve these estimates and generalize them to $\ell\geq 1$.

Our results establish the Laughlin functions \eqref{def:Lau} as exact ground states of magnetic Hamiltonians with interactions through a rigorous limit procedure. We emphasize, however, that this procedure requires a strong interaction potential of short range. In theoretical studies of Bose-Einstein condensation in dilute atomic gases \cite{PS,LSSY}, the interactions between atoms and molecules are often modeled by strong repulsive potentials, even as hard spheres, of range much smaller than the mean particle distance. Such models are prime examples to which our results apply, the $\ell=0$ case for bosons treated in \cite{LS} being the simplest one. The role of the magnetic field may here be taken over by the angular velocity of a system in rapid rotation \cite{C,RSY}.

In quantum Hall physics for electrons, on the other hand, the dominant interaction is the Coulomb interaction between the particles (in addition to external potentials modeling traps and impurities) and is thus of a different character than in the examples just mentioned. Numerical studies indicate that Laughlin states may have large overlap with true ground states in quantum Hall settings (see, e.g.,  \cite{PG}, Sect. 8.7), and,  in fact, Haldane's expansion \eqref{Hpp} applied to a Coulomb potential produces terms that decrease rapidly with increasing $\ell$. So far, no rigorous justification for the pseudo-potential description in the Coulomb case is known, however.


\section{Model and main results}


On $L^2(\R^2)$, define the operator $h\geq 0$ as 
\beq\label{def:h}
h = ( -\rmi \nabla - x^\perp)^2 - 2
\eeq
where $x^\perp = (-x^{(2)},x^{(1)})$ for $x=(x^{(1)},x^{(2)})\in \R^2$. It is simply the magnetic Laplacian for magnetic field $B=2$, with its ground state energy shifted to zero, having spectrum $4\N \cup \{0\}$. The particle mass has been set to $1/2$ and Planck's constant to 1. For $v\geq 0$ radial and of compact support, and $N\geq 2$, define the $N$-particle Hamiltonian
\beq\label{hamiltonian}
H_a = H^{(0)}+ \sum_{1\leq i<j\leq N} v_a ( |x_i-x_j| )
\eeq
on $\Hh= L^2(\R^{2N})$, where $H^{(0)}=\sum_{i=1}^N h_i$ is the $N$-particle magnetic kinetic energy operator and 
\beq
v_a(r) = a^{-2} v(r/a)
\eeq
for $a>0$. This scaling appears naturally in our choice of units, reflecting the fact that $-\nabla^2$ scales like the square of an inverse length. For short-range interactions, the scattering length is the natural parameter measuring their strength, and the scattering length of $v_a$ equals $a$ times the one of $v$. We are interested in the regime where $a$ is much smaller than the magnetic length, which is $O(1)$ in our units, i.e., we consider the case $a\ll 1$. In particular, the strength $a^{-2}$ of the interaction is much larger than the energy gap between Landau levels.\footnote{Alternatively, we could consider the system for a fixed $v$ (corresponding to the choice $a=1$)  
and subject to a magnetic field $B>0$. Via a rescaling of lengths, the resulting Hamiltonian is then unitarily equivalent to $B/2$ times  $H_{\sqrt{B/2}}$. In other words, the parameter $a$ in \eqref{hamiltonian} has the physical interpretation of the ratio of the scattering length of  the interaction potential to the magnetic length.}

We assume that $v$ is a non-negative measurable function of compact support, but we don't need to assume that $v$ is integrable; in particular, it is allowed to have a hard core, i.e., be infinite on a set of positive measure. Functions in the domain of the Hamiltonian then vanish on the corresponding set in configuration space.
It is not necessary to restrict to symmetric or anti-symmetric functions, our analysis is valid on the whole Hilbert space $\Hh= L^2(\R^{2N})$ and applies equally to bosons and fermions. 
 
 We introduce the sequence of closed subspaces
\beq
\Hh \supset \B_0 \supset \B_1 \supset \dots
\eeq
where $\B_\ell$ for $\ell \geq 0$  consists of $\Psi \in \Hh$ of the form
\beq
 \Psi(x_1,\dots,x_N) = \rme^{-\tfrac 12 \sum_{i=1}^N |x_i|^2}  \varphi(z_1,\dots,z_N) \prod_{i<j} (z_i - z_j)^\ell 
\eeq
 with $ \varphi : \C^N\to \C^N$ analytic. Here we identify $z_j \in \C$ with $x_j = (x^{(1)}_j, x^{(2)}_j) \in \R^2$ via  $z_j = x^{(1)}_j + \rmi x^{(2)}_j$. Note that $\B_0$ coincides with the LLL, i.e., the kernel of $H^{(0)}$. We also note  that for large $N$ all normalized wave functions in $\B_\ell$ have a remarkable incompressibility property: their one-particle density, suitably averaged, is everywhere bounded above by $(\pi \ell)^{-1}$ \cite{LRY,OR,R}.

On the space $\B_\ell$, we define the operators
\beq
\hf_\ell = \sum_{i<j} \D^{(\ell)}_{ij}
\eeq
where $\D^{(\ell)}$ is for $\ell\geq 0$ defined via the quadratic form for a two-particle wave function $\Psi(x_1,x_2) = \rme^{-\tfrac 12 (|x_1|^2+|x_2|^2)} \varphi(z_1,z_2) (z_1-z_2)^\ell$ as 
\beq\label{def:dell}
\langle \Psi | \D^{(\ell)} \Psi\rangle =    \int_{\R^2} \rme^{-2|x|^2} | \phi(z,z)|^2 dx
\eeq
and  $\D^{(\ell)}_{ij}$ acts on an $N$-particle wave function in $\B_\ell$ like $\D^{(\ell)}$ w.r.t.\ the variables $z_i$, $z_j$ if the others are fixed. Note that $\D^{(\ell)}_{ij}$ is a bounded operator; in fact, by comparing expectation values and using  \eqref{def:phiell}, one sees that  
\beq 
\D^{(\ell)}_{ij}=(\pi \ell!)^{-1}\mathfrak P_{ij}^{(\ell)}
\eeq
on $\B_\ell$,
with the previously introduced projection $\mathfrak P_{ij}^{(\ell)}$ on states with relative angular momentum $\ell$ for a pair $ij$. Hence $\hf_\ell$ is equal to  
\eqref{def:Pell} on $\B_\ell$, up to the factor $(\pi \ell!)^{-1}$. Moreover,  the kernel of $\hf_\ell$ coincides with $\B_{\ell+1}$, the domain of $\hf_{\ell+1}$.

Next we define the relevant scattering parameters in arbitrary angular momentum channels, generalizing the approach to the $s$-wave scattering length  in \cite[App.~C]{LSSY} and \cite{LY2}.
For $\ell \in \N$, $\ell \geq 1$, define $b_\ell$ via the variational principle
\begin{equation}\label{def:bell}
b_\ell = \frac 1{4 \pi \ell }  \min \left\{ \int_{\R^2} |x|^{2\ell} \left( |\nabla f(x)|^2 + \tfrac 12 v(|x|) |f(x)|^2 \right) dx \, : \, \lim_{|x|\to\infty} f(x) = 1\right\}.
\end{equation}
We denote the minimizer by $f_\ell$. It is radially symmetric, satisfies $0\leq f_\ell \leq 1$, and  $f_\ell(|x|) = 1 - b_\ell/|x|^{2\ell}$ for $x$ outside the support of $v$, i.e., $|x|> R_0$. In particular, $b_\ell = R_0^{2\ell}$ for hard discs. The variational equation (zero-energy scattering equation in the $\ell$ channel) for the minimizer reads, with $r=|x|$,
\beq \label{def:vareq}
- f_\ell^{\prime\prime}(r)-(2\ell+1)r^{-1}f_\ell^\prime(r)+\half v(r)f_\ell(r)=0.
\eeq
For $\ell=0$ the zero energy scattering solution is not bounded but rather grows logarithmically at infinity. We take $R>R_0$ and define the $s$-wave scattering length $b_0$ as in \cite{LY2} by
\begin{equation}\label{def:b0}
\frac {4\pi} {\ln(R^2/b_0^2)} =   \min \left\{ \int_{\R^2}  \left( |\nabla f(x)|^2 + \tfrac 12 v(|x|) |f(x)|^2 \right) dx \, : \,  f(x) = 1 \hbox{ for } |x|\geq R\right\}.
\end{equation}
As shown in \cite{LY2}, $b_0$ is independent of $R$; for hard discs $b_0=R_0$.

If $v$ is replaced by $v_a$ then, by a change of variables in the integrals \eqref{def:bell} and \eqref{def:b0}, one sees that $b_\ell$ is replaced by $a^{2\ell}b_\ell$ for $\ell\geq 1$ and $b_0$ by $ab_0$.

\bigskip

With these definitions, our main result can be formulated as follows.

\begin{theorem}\label{thm:main}
For any $\ell\geq 1$, the operator $a^{-2\ell} H_a$ converges to $ 8\pi \ell b_\ell \hf_\ell$ in strong resolvent sense as $a\to 0$, i.e., for any $\mu > 0$ and $\psi \in \Hh= L^2(\R^{2N})$ 
\beq\label{2.9}
\lim_{a\to 0} (\mu + a^{-2\ell} H_a )^{-1} \psi  =  (\mu+ 8\pi \ell b_\ell \hf_\ell )^{-1} P_\ell \psi
\eeq
strongly in $L^2(\R^{2N})$, where $P_\ell$ denotes the projection onto  $\B_\ell \subset \Hh$. Moreover, for $\ell=0$, $\ln (1/a^2)  H_a$ converges in the same sense to $8\pi \hf_0$. 
\end{theorem}

\begin{remark}
In Theorem~\ref{thm:main} we assume that the interaction potential $v$ is not identically zero. If it is, we have $b_\ell = 0$ for all $\ell$, and \eqref{2.9} trivially holds with $P_\ell$ replaced by $P_0$ in this case. 
\end{remark}

\begin{remark} If one introduces a coupling parameter, i.e., replaces $v$ by $\lambda v$ for $\lambda>0$, one can explore, in addtion, the regimes of weak and strong coupling. In case $v$ is suitably regular and $\lambda\ll 1$, the parameters $b_\ell$ are to leading order given by their Born approximation, $b_\ell \approx (8\pi \ell)^{-1} \lambda \int r^{2\ell} v$, corresponding to the choice $f\equiv 1$ in \eqref{def:bell}. That is, for weak potentials, one recovers the moments of $v$ as the pre-factors of the pseudo-potentials, as predicted by first-order perturbation theory. In the limit of strong coupling $\lambda\gg 1$, on the other hand, one obtains the scattering parameters of hard spheres, solely determined by their diameter.
\end{remark}

\begin{figure}[h]
\centering
\includegraphics[width=10cm]{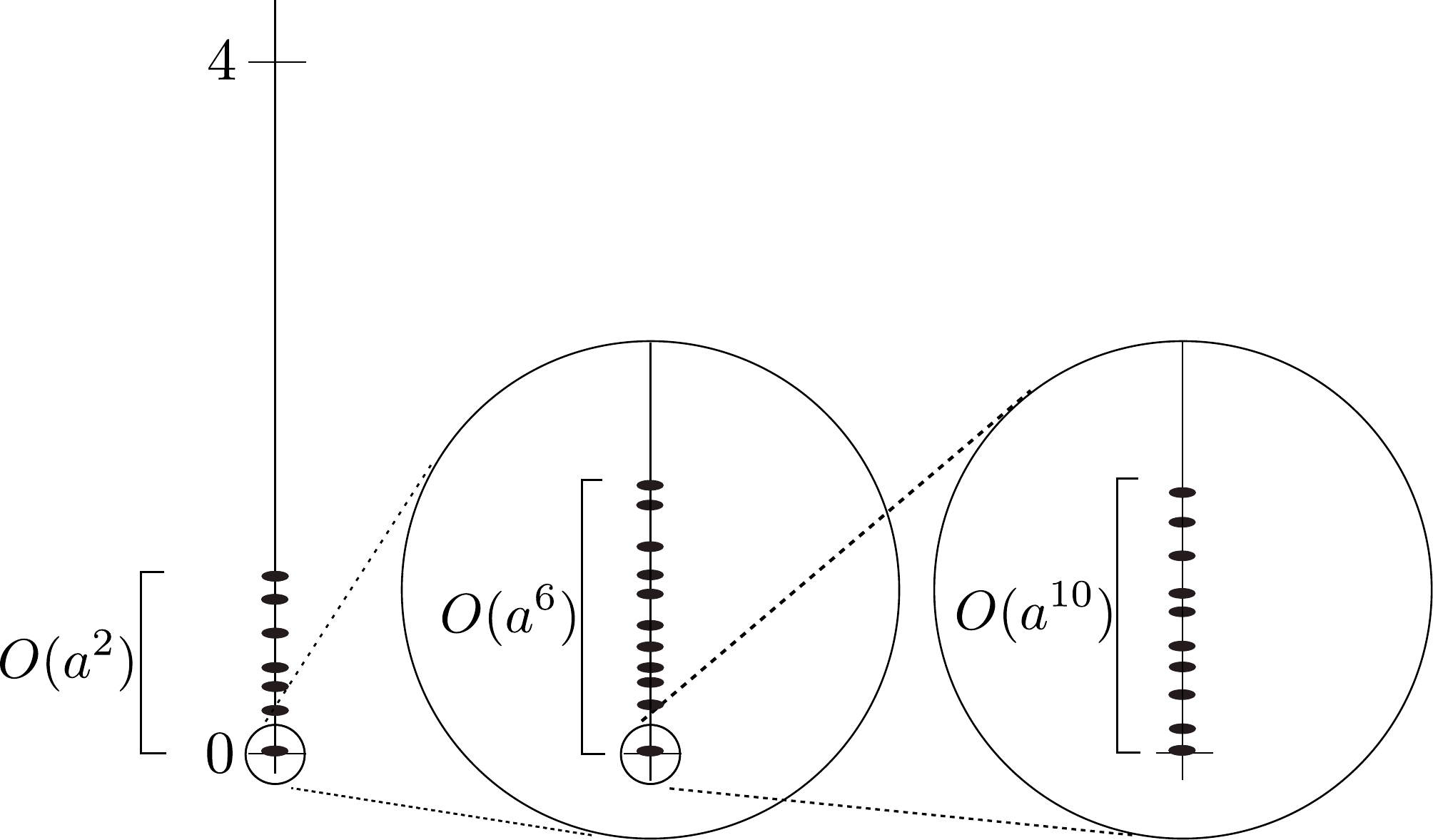}
\caption{Sketch of the  spectrum in the fermionic case. States with energy of order $a^2$ are  described by the effective Hamiltonian $\hf_{1}$ on the LLL. Zooming in on its kernel, one finds $\hf_{3}$ as an effective Hamiltonian, describing states with energy of order $a^6$, and so forth.\label{fig1}}
\end{figure}

The result in Theorem~\ref{thm:main} can be interpreted as follows. States in $\Hh$ with energy of order $a^{2\ell}$ are, for small $a$, necessarily close to states in $\B_\ell$, and are described by an effective Hamiltonian $\hf_\ell$. On the kernel of $\hf_\ell$, one can zoom in further by looking at energies of order $a^{2\ell'}$ for some $\ell'>\ell$, and find a new effective Hamiltonian $\hf_{\ell'}$. In the limit $a\to 0$, one thus obtains an infinite cascade of effective Hamiltonians in the corresponding energy windows.  See Fig.~\ref{fig1} for an illustration.

Theorem~\ref{thm:main} readily implies that, for any hermitian bounded operator $K$ on $\Hh$ that is independent of $a$, 
\begin{equation}\label{2:10}
  a^{-2\ell} H_a  + K  \  \xrightarrow{a\to 0}  \ 8\pi \ell b_\ell \hf_\ell + P_\ell K P_\ell
\end{equation}
in strong resolvent sense, and this generalizes to suitable unbounded $K$. One could  consider an additional interaction potential, for instance of Coulomb type, in which case $P_\ell K P_\ell$ is a linear combination of the operators $\sum_{i\neq j}\mathfrak P^{(\ell')}_{ij}$ as in  \eqref{Hpp}, but with the sum restricted to angular momenta  $\ell'\geq \ell$.  
If one adds a confining potential, it is not difficult to show 
that the convergence in \eqref{2:10} holds actually in the norm-resolvent sense, i.e., the operator norm of the difference between the resolvents of the left and right side of \eqref{2:10} tends to zero as $a\to 0$. In fact, the resolvents are in this case compact operators and $( \mu + a^{-2\ell} H_a +K)^{-1} \leq (\mu + H^{(0)} + K)^{-1}$ for any $a\leq 1$ and $\mu> - \infspec K$, where we used the assumption that $v$ is non-negative. One readily checks that strong- and norm-resolvent convergence are equivalent for  a sequence of non-negative operators that are dominated by a fixed compact operator. 

A convenient choice is the harmonic oscillator potential $K= \sum_{i=1}^N |x_i|^2$, which acts as $P_\ell K P_\ell =  (N+L)P_\ell$, with $L = \sum_{i=1}^N (z_i \partial_{z_i}-\bar z_i \partial_{\bar z_i}) =  \sum_{i=1}^N z_i \partial_{z_i}$ the total angular momentum operator in the LLL. Indeed, $\ell\delta_{\ell m}=\langle \varphi_\ell, L \varphi_m\rangle=\langle \varphi_\ell,(|z|^2-1)\varphi_m\rangle$ using \eqref{def:phiell}. In particular, for any $\lambda>0$ and $\ell \geq 1$ we have 
\begin{equation}\label{eq1}
  a^{-2\ell} H_a + \lambda \sum_{i=1}^N |x_i|^2  \  \xrightarrow{a\to 0}  \ 8\pi \ell b_\ell \hf_\ell + \lambda (N+L)P_\ell
\end{equation}
in {\em norm}-resolvent sense. This extends to $\ell =0$ in the same way as explained in Theorem~\ref{thm:main}. 

The right side of \eqref{eq1} is the sum of two commuting operators on $\B_\ell$, which we write for short as
\beq\gamma \hf_\ell+\lambda \mathcal L\label{sum}\eeq with
$\mathcal L=(N+L)P_\ell$ and $\gamma,\lambda>0$, and we consider the joint spectrum of the commuting operators $\hf_\ell$ and $\mathcal L$.  For a fixed eigenvalue of $\mathcal L$ we have a finite dimensional space on which $\hf_\ell$ has nonnegative eigenvalues,  and the situation is analogous to the discussion of the {\em Yrast curve} (see \cite[Fig.~1]{LS} for a sketch). This curve is the (convex hull of the)  boundary of the joint spectrum. By varying the ratio $\lambda/\gamma$  we can adjust the value(s) of the angular momentum where the energy is minimal and thereby change the ground state(s) of \eqref{sum}. Indeed, the ground state is determined by the point(s) where a line with slope $-\lambda/\gamma$ touches the joint spectrum.  For fixed $\gamma$ and $\lambda$ small enough (depending on the spectral gap of $\hf_\ell$) the unique ground state equals the Laughlin state $\Psi^{\rm L}_{\ell+1}$, where $\hf_\ell$ has eigenvalue 0. It is separated from other states with the same or less angular momentum by a spectral gap since the state space is finite dimensional, but the size of the gap might {\it a-priori} depend on the particle number $N$. It is an important, but still unproved, conjecture in FQHE physics that this gap has a strictly positive lower bound independent of $N$ (see, e.g., \cite{R}).  For
 $\lambda$ large enough the unique ground state of \eqref{sum}  is the Laughlin wave function $\Psi^{\rm L}_\ell$, where $\hf_\ell$ is strictly positive but $\mathcal L$ takes its minimal value in $\B_\ell$.

 The norm-resolvent convergence in \eqref{eq1} implies convergence of eigenvalues and corresponding eigenvectors, and  hence we can conclude the following corollary:

\begin{corollary}\label{cor1}
Fix $\ell\geq 1$. For $\lambda>0$ small enough, the ground state of  $H_a + a^{2\ell} \lambda \sum_{i=1}^N |x_i|^2$ converges in $L^2(\R^{2N})$ to $\Psi^{\rm L}_{\ell+1}$ as $a\to 0$. For $\ell=0$ this holds with $a^{2\ell}$ replaced by $(\ln(1/a^2))^{-1}$.
\end{corollary}

These results hold on the whole Hilbert space $\Hh=L^2(\R^{2N})$ without symmetry constraints. When restricting to the bosonic and fermionic subspaces, respectively, we have $\B_{\ell-1} = \B_{\ell}$ for $\ell$ even (bosons) or $\ell$ odd (fermions), hence it is natural to restrict to such $\ell$ depending on the symmetry. The unique ground state of the right side of \eqref{eq1} is now $\Psi^{\rm L}_{\ell +2}$ for small $\lambda$. In particular, Corollary~\ref{cor1} holds with $\Psi^{\rm L}_{\ell+1}$ replaced by $\Psi^{\rm L}_{\ell+2}$ in the bosonic subspace for even $\ell$, and in the fermionic subspace for odd $\ell$.

\begin{remark}
Our results can be extended  to a system in three spatial dimensions with an additional confinement in the third direction. This will be detailed in Section~\ref{sec:3d}, thereby generalizing a corresponding result for the special case $\ell=0$ in \cite{LS}.
\end{remark}

\section{Proof of Theorem~\ref{thm:main}}

\subsection{Gamma convergence}

It is well known that strong resolvent convergence of  operators  is equivalent to $\Gamma$-convergence of the corresponding quadratic forms in both weak and strong topologies  (see \cite[Sec.~13]{masi}). Hence we find it convenient to reformulate Theorem~\ref{thm:main} in terms of $\Gamma$-convergence.

We define the functions $F_a : \Hh \to [0,\infty]$ as 
\beq
F_a(\Psi) = \| \sqrt{H_a} \Psi \|^2
\eeq
(with the understanding that $F_a$ equals $+\infty$ if $\Psi$ is not in the form domain of $H_a$). Moreover, for $\ell\geq 0$, define $G^{(\ell)}:\Hh \to [0,+\infty]$ as
\beq
G^{(\ell)}(\Psi) = \begin{cases} \langle \Psi | \hf_\ell \Psi\rangle & \text{if $\Psi \in \B_\ell$} \\ +\infty & \text{otherwise.} \end{cases}
\eeq
Theorem~\ref{thm:main} is an immediate consequence of the following Proposition.

\begin{proposition}\label{prop:main}
For any $\ell\geq 1$, the function $a^{-2\ell} F_a$ $\Gamma$-converges to $8\pi \ell b_\ell G^{(\ell)}$ as $a\to 0$ both in the strong and weak topology on $\Hh$. That is, for $\ell \geq 1$, 
\begin{enumerate}
\item for any sequence $\{a_n\}_{n\in \N}$ of positive numbers with $\lim_{n\to \infty} a_n = 0$, and any sequence $\{\Psi_n\}_{n\in \N}$ in $\Hh$ with $\Psi_n \rightharpoonup \Psi \in \Hh$ as $n\to \infty$, 
\begin{equation}\label{eq:thmlb}
\liminf_{n\to \infty} a_n^{-2\ell} F_{a_n} ( \Psi_n) \geq  8\pi \ell b_\ell G^{(\ell)}(\Psi)\,.
\end{equation}
\item for any $\Psi \in \Hh$ and any sequence $\{a_n\}_{n\in \N}$ of positive numbers with $\lim_{n\to \infty} a_n = 0$, 
 there exists a sequence $\{\Psi_n\}_{n\in \N}$ in $\Hh$ with $\Psi_n \to \Psi \in \Hh$ as $n\to \infty$ such that
\begin{equation}\label{eq:thmub}
\lim_{n\to \infty} a_n^{-2\ell} F_{a_n} ( \Psi_n) =  8\pi \ell b_\ell G^{(\ell)}(\Psi)\,.
\end{equation}
\end{enumerate}
For $\ell=0$, the same holds with $a^{-2\ell}$ replaced by $\ln(1/a^2)$, and $\ell b_\ell$ replaced by $1$.
\end{proposition}

Note that \eqref{eq:thmlb} holds for all weakly converging sequences, while the convergence in (2) is strong, hence indeed there is $\Gamma$-convergence in both topologies. 
The proof of Proposition~\ref{prop:main} will be given in the remainder of this section. We discuss the case $\ell\geq 1$ in detail, and indicate the modifications for $\ell=0$ in the last Subsection, \ref{ss3.5}.

\subsection{Dyson Lemma}

We start with the following {\em Dyson Lemma}, so named because Dyson  proved a first estimate of this type in 1957 \cite{Dys} to obtain a lower bound on the energy of a hard core Bose gas. Subsequently this was extended in several ways, see \cite{LY1, LY2, LSS} and, for the present context,  in particular \cite{LS}.  Lemma~4 in \cite{LS} is stated and proved for $\ell=0$, but since in that paper only the three-dimensional case was  considered, we shall need an extra discussion for $\ell=0$ here, cf. Sec.~\ref{ss3.5}.

We again identify $x=(x^{(1)},x^{(2)}) \in \R^2$ with $z = x^{(1)} + \rmi x^{(2)} \in \C$, and write $\partial_{\bar z} = \frac 12( \partial_{x^{(1)}} +\mathrm  i \partial_{x^{(2)}})$, $\partial_{z} = \frac 12(\partial_{x^{(1)}} - \rmi \partial_{x^{(2)}})$. 

\begin{lemma}
For any $\ell \geq 1$ and $R>a R_0$, and any $y\in\C$, we have
\begin{align}\nonumber
 &\int_{|z-y| < R}  \rme^{-|z|^2} \left( 4 |\partial_{\bar z} \phi(x)|^2 + \tfrac 12 v_a(|z-y|) |\phi(x)|^2 \right) dx 
 \\ & \geq 4\pi \ell b_\ell a^{2\ell}  \rme^{|y|^2 -R^2 }  \left|  \frac 1{2\pi \rmi} \oint_{|z-y|=R}  \rme^{-z \bar y}\frac{  \phi(x)}{(z-y)^{\ell +1}} dz \right|^2 \label{eq:dyson}
\end{align}
where $dz$ stands for the complex line element.
\end{lemma}

Note that if $\phi$ is analytic and vanishes like $\kappa (z-y)^{\ell}$ as $z\to y$ for some $\kappa \in \C$, 
the right side is proportional to $|\kappa|^2 \rme^{ -|y|^2 }$.

\begin{proof}
For $R>a R_0$, consider the expression
\beq
A = \int_{|z|<R} \bar z^\ell \left[ 4 \left( \partial_z f(x) \right) \left( \partial_{\bar z} \phi(x) \right) + \tfrac 12 v_a(|z|) f(x) \phi(x) \right] dx 
\eeq
where $f(x) = f_\ell(x/a)$ with $f_\ell$ the minimizer in \eqref{def:bell}. 
The Cauchy--Schwarz inequality and the fact that $|\partial_z f| = \frac 12 |\nabla f|$ (since $f$ is real) imply that 
\begin{align}\nonumber
|A|^2 &  \leq \int_{|x| < R} |x|^{2\ell} \left( |\nabla f(x)|^2 + \tfrac 12 v_a( |x|) |f (x)|^2 \right) dx 
\\ & \quad \times \int_{|x| < R}  \left( 4 |\partial_{\bar z} \phi(x)|^2 + \tfrac 12 v_a(|x|) |\phi(x)|^2 \right) dx.
\end{align}
The term in the first line is bounded above by $4\pi \ell  b_\ell a^{2\ell}$, hence 
\beq
 \int_{|x| < R}  \left( 4 |\partial_{\bar z} \phi(x)|^2 + \tfrac 12 v_a(|x|) |\phi(x)|^2 \right) dx \geq \frac{|A|^2}{4\pi \ell b_\ell a^{2\ell}}.
 \eeq

Integrating by parts and using the variational equation \eqref{def:vareq} for $f$ (and the fact that $f$ is radial), we also have
\begin{align}\nonumber
A & = \int_{|z|<R} |z|^{2\ell} \left[  4 \left( \partial_z f(x) \right) \left( \partial_{\bar z} z^{-\ell}\phi(x) \right) +\tfrac 12 v_a( |z|) f(x) z^{-\ell} \phi(x) \right] dx 
\\ & 
= 4\pi  \ell b_\ell a^{2\ell} \frac 1{2\pi \rmi} \oint_{|z|=R} \frac{ \phi(x)}{z^{\ell +1}} dz.
\end{align}
We conclude that
\beq
 \int_{|x| < R}  \left( 4 |\partial_{\bar z} \phi(x)|^2 + \tfrac 12 v_a(|x|) |\phi(x)|^2 \right) dx \geq 4\pi \ell b_\ell a^{2 \ell}  \left|  \frac 1{2\pi\rmi} \oint_{|z|=R} \frac{ \phi(x)}{z^{\ell +1}} dz \right|^2.
 \eeq
To bring this into the desired form, we bound for $y\in \C$ 
\begin{align}\nonumber 
 &\int_{|x| < R}  \rme^{-|z+y|^2} \left( 4 |\partial_{\bar z} \phi(x)|^2 + \tfrac 12 v_a(|x|) |\phi(x)|^2 \right) dx 
 \\ & = \rme^{-|y|^2}  \int_{|x| < R}  \rme^{-|z|^2} \left( 4 |\partial_{\bar z} \rme^{-z \bar y}\phi(x)|^2 +\tfrac 12 v_a( |x|) | \rme^{-z \bar y}\phi(x)|^2 \right) dx \nonumber
 \\ & \geq \rme^{-|y|^2 -R^2 }  \int_{|x| < R}   \left( 4 |\partial_{\bar z} \rme^{-z \bar y}\phi(x)|^2 +\tfrac 12 v_a( |x|) | \rme^{-z \bar y}\phi(x)|^2 \right) dx \nonumber
 \\ & \geq 4\pi \ell b_\ell a^{2\ell} \rme^{-|y|^2 -R^2 }  \left|  \frac 1{2\pi \rmi} \oint_{|z|=R}  \rme^{-z \bar y}\frac{  \phi(x)}{z^{\ell +1}} dz \right|^2.
\end{align}
In particular, changing variables from $z$ to $z-y$, Eq.~\eqref{eq:dyson} follows.
\end{proof}

By averaging the bound \eqref{eq:dyson} over $R$, we obtain as an immediate corollary
\begin{align}\nonumber
 &\int_{|z-y| < R}  \rme^{-|z|^2} \left(4 |\partial_{\bar z} \phi(x)|^2 + \tfrac 12 v_a(|z-y|) |\phi(x)|^2 \right) dx 
 \\ & \geq 4\pi \ell b_\ell a^{2\ell}  \rme^{|y|^2  }  \int_0^\infty \rme^{-r^2} \rho(r) \left|  \frac 1{2\pi \rmi} \oint_{|z-y|=r}  \rme^{-z \bar y}\frac{  \phi(x)}{(z-y)^{\ell +1}} dz \right|^2 dr \label{eq:dyson2}
\end{align}
for any non-negative function $\rho$ supported on $[a R_0, R]$ with $\int \rho = 1$.  
We shall choose $\rho$ bounded, supported on $[R/2,R]$, and independent of $a$; an explicit choice is $\rho(r)=2/R$ for $R/2<r<R$, and $0$ otherwise. As long as $a < R/(2R_0)$, \eqref{eq:dyson2} holds for this choice of $\rho$. 

\subsection{Lower bound}\label{ss:lb}

We now turn to the proof of part (1) of Prop.~\ref{prop:main} and establish  the lower bound \eqref{eq:thmlb}. We start by noting that, for any $\Psi \in L^2(\R^2)$ of the form $\Psi(x) = \rme^{-|x|^2/2} \phi(x)$, we have the representation 
\begin{equation}\label{eq:rep}
\langle \Psi | h \Psi \rangle = 4\int_{\R^2} \rme^{-|x|^2} \left| \partial_{\bar z} \phi(x) \right|^2 dx
\end{equation}
where we denote $\partial_{\bar z} =\half( \partial_{x^{(1)}} + \rmi \partial_{x^{(2)}})$ as above. 
This representation, in combination with \eqref{eq:dyson2},  implies the lower bound
\begin{align}\nonumber
a^{-2\ell} F_a(\Psi) & \geq 4\pi \ell b_\ell   \sum_{i \neq j}^N  \int_{\R^{2(N-1)}} \rme^{- \sum_{k, k\neq i,j }^N |x_k|^2} \chi_{i,R}(x_1,\dots, \, \not \! x_i, \dots, x_N) \\ & \quad \times \int_0^\infty \rme^{-r^2} \rho(r) \left|  \frac 1{2\pi \rmi} \oint_{|z_i-z_j|=r}  \rme^{-z_i \bar z_j }\frac{  \phi(x_1,\dots,x_N)}{(z_i-z_j)^{\ell +1}} dz_i \right|^2 dr\, \prod_{j, j\neq i}^N dx_j \label{lb1}
\end{align}
for $\Psi(x_1,\dots,x_N) = \rme^{-\tfrac 12 \sum_{i=1}^N |x_i|^2} \phi(x_1,\dots,x_N)$, where
\beq
 \chi_{i,R}(x_1,\dots, \, \not \! x_i, \dots, x_N)= \prod_{j<k, j,k\neq i} \theta( |x_j - x_k| - 2R)
\eeq
restricts the integration to the set where $|x_j-x_k|\geq 2R$ for all $j,k\neq i$. 

We claim that the quadratic form defined on the right side of \eqref{lb1} is bounded. In fact, by a simple Cauchy--Schwarz inequality, using that $\Re z_i \bar z_j = \frac 12 (|z_i|^2 + |z_j|^2- |z_i-z_j|^2 )$, 
\begin{align}\nonumber
&  \left|  \frac 1{2\pi \rmi} \oint_{|z_i-z_j|=r}  \rme^{-z_i \bar z_j }\frac{  \phi(x_1,\dots,x_N)}{(z_i-z_j)^{\ell +1}} dz_i \right|^2  \\ & \leq 
\frac { \rme^{r^2} }{2\pi r^{2\ell+1} } \oint_{|z_i-z_j|=r}  \rme^{-|x_i|^2 - |x_j|^2} |\phi(x_1,\dots,x_N)|^2 | dz_i| \,,
\end{align}
which after integration over $r$ and $x_j$ implies that
\begin{align}\nonumber
& \int_0^\infty \rme^{-r^2} \rho(r) \int_{\R^2}  \left|  \frac 1{2\pi \rmi} \oint_{|z_i-z_j|=r}  \rme^{-z_i \bar z_j }\frac{  \phi(x_1,\dots,x_N)}{(z_i-z_j)^{\ell +1}} dz_i \right|^2 dr \, dx_j \\ & \leq \frac 1{2\pi} \int_{\R^4} \rme^{-|x_i|^2 - |x_j|^2} |\phi(x_1,\dots,x_N)|^2 \frac {\rho(|x_i-x_j|)}{|x_i - x_j|^{2\ell +1}} dx_i \, dx_j.
\end{align}
 In particular, as a bounded quadratic form,  the right side of \eqref{lb1} is weakly lower semicontinuous. 

Now given a sequence $a_n\to 0$ and $\Psi_n \rightharpoonup \Psi$, we obtain from the bound above and weak lower semicontinuity 
\begin{align}\nonumber
& \liminf_{n\to \infty} 
 a_n^{-2\ell} F_{a_n}(\Psi_n) \\ \nonumber & \geq 4\pi \ell b_\ell   \sum_{i \neq j}^N  \int_{\R^{2(N-1)}} \rme^{- \sum_{k, k\neq i,j }^N |x_k|^2} \chi_{i,R}(x_1,\dots, \, \not \! x_i, \dots, x_N) \\ & \quad \times \int_0^\infty \rme^{-r^2} \rho(r) \left|  \frac 1{2\pi \rmi} \oint_{|z_i-z_j|=r}  \rme^{-z_i \bar z_j }\frac{  \phi(x_1,\dots,x_N)}{(z_i-z_j)^{\ell +1}} dz_i \right|^2 dr\, \prod_{j, j\neq i}^N dx_j \label{lb2}
\end{align}
where $\Psi(x_1,\dots,x_N) = \rme^{-\tfrac 12 \sum_{i=1}^N |x_i|^2} \phi(x_1,\dots,x_N)$. Consider first the case when $\Psi \in \B_\ell$. Then $\phi$ is analytic and of the form $\phi(x_1,\dots,x_N) = \tilde \phi(z_1,\dots,z_N) \prod_{i<j} (z_i-z_j)^\ell$ for some analytic $\tilde \phi$. Writing $\phi(x_1,\dots,x_N) =  \xi_{ij}(z_1,\dots,z_N)  (z_i-z_j)^\ell$, we have
\beq
\frac 1{2\pi \rmi} \oint_{|z_i-z_j|=r}  \rme^{-z_i \bar z_j }\frac{  \phi(x_1,\dots,x_N)}{(z_i-z_j)^{\ell +1}} dz_i = \rme^{-|z_j|^2} \xi_{ij}(z_1,\dots,z_j, \dots , z_j, \dots, z_N)
\eeq
in this case. Hence the right side of \eqref{lb2} equals
\begin{align}\nonumber
& 4\pi \ell b_\ell \int_0^\infty \rme^{-r^2} \rho(r) dr\sum_{i \neq j}^N  \int_{\R^{2(N-1)}} \rme^{- \sum_{k, k\neq i,j }^N |x_k|^2} \rme^{-2 |z_j|^2}\chi_{i,R}(x_1,\dots, \, \not \! x_i, \dots, x_N) \\ & \qquad \quad \times |\xi_{ij}(z_1,\dots,z_j, \dots , z_j, \dots, z_N)|^2 \prod_{j, j\neq i}^N dx_j . \label{lb3}
\end{align}
We have $\int \rme^{-r^2} \rho(r) dr \geq \rme^{-R^2}$, which goes to $1$ as $R\to 0$. Moreover, by dominated convergence, we can replace $\chi_{i,R}$ by $1$ in the limit $R\to 0$, and conclude that 
\beq
\lim_{R\to 0} \eqref{lb3} = 
4\pi \ell b_\ell  \sum_{i\neq j}  \langle \Psi | \D^{(\ell)}_{ij} \Psi\rangle = 8\pi \ell b_\ell \, G^{(\ell)}(\Psi) .
\eeq
Since $R>0$ was arbitrary, this yields  the desired lower bound.

Next, consider the case when $\Psi \in \B_{\ell'}$ for some $\ell' < \ell$, but $\Psi \not\in \B_{\ell'+1}$ (and hence, in particular, $\Psi \not\in \B_\ell$). In this case, we can apply the bound \eqref{lb2} with $\ell$ replaced by $\ell'$, to conclude that $\liminf_{n\to \infty} a_n^{-2\ell} F_{a_n} (\Psi) = + \infty$, as desired. Here we use the fact that the kernel of $\hf_{\ell'}$ equals $\B_{\ell'+1}$, hence $\Psi$ is not in the kernel. 

Finally, consider the case $\Psi\not \in \B_0$. Then we can simply drop the interaction for a lower bound, and conclude that $\liminf_{n\to\infty} F_{a_n}(\Psi_n) \geq \| \sqrt{H^{(0)}} \Psi \|^2>0$. In particular, $\liminf_{n\to\infty} a_n^{-2\ell} F_{a_n}(\Psi_n) = + \infty$ for any $\ell\geq 1$. This concludes the proof of the lower bound for $\ell\geq 1$.

\subsection{Upper Bound}

We shall now prove part (2) of Prop.~\ref{prop:main}. Given the lower bound \eqref{eq:thmlb} we already established, we only need to prove  \eqref{eq:thmub} as an upper bound. It clearly suffices to consider the case $\Psi \in \B_\ell$. In the opposite case $\Psi \not \in \B_\ell$, the energy tends to infinity and can simply take $\Psi_n = \Psi$ for all $n$, and use  the lower bound \eqref{eq:thmlb}.

For $\Psi \in \B_\ell$, we consider the sequence 
\beq
\Psi_n(x_1,\dots,x_N) = \Psi(x_1,\dots,x_N) \prod_{i<j} f(x_i - x_j)
\eeq
where $f   (x) = f_{n,\ell}(x)=f_\ell(x/a_n)$, with $f_\ell$ the minimizer in \eqref{def:bell}. Since $f$ is bounded and converges pointwise to $1$ as $a_n\to 0$, $\Psi_n\to \Psi$ by dominated convergence. Using the representation \eqref{eq:rep}, we have
\beq
F_{a_n}(\Psi_n) = \int_{\R^{2N}} |\Psi(x_1,\dots,x_N)|^2
\left[ 4 \sum_{i=1}^N | \partial_{\bar z_i} S|^2 + \half \sum_{i<j}  v_{a_n}( |x_i-x_j |) | S |^2 \right] \prod_{j=1}^N dx_j
\eeq
for $S(x_1,\dots,x_N) = \prod_{i<j} f(x_i-x_j)$. Since $S$ is real, $ | \partial_{\bar z_i} S| =  \frac 12 | \nabla_i S|$. Moreover, since $0\leq f\leq 1$, we can bound $|S|^2 \leq f(x_i-x_j)^2$ for any pair $i\neq j$, as well as 
\beq
\sum_{i=1}^N |\nabla S|^2 \leq \sum_{i\neq j} |\nabla f(x_i-x_j)|^2 + \sum_{i\neq j\neq k} |\nabla f(x_i-x_j)| | \nabla f(x_k -x _j)| .
\eeq
We thus have to bound terms of the form
\beq
{\rm I}=\int_{\R^{2N}} |\Psi(x_1,\dots,x_N)|^2\left\{|\nabla f(x_i-x_j)|^2+\half v_{a_n}(|x_i-x_j |)|f(x_i-x_j)|^2\right\}\prod_{l=1}^N dx_l
\eeq
and
\beq
{\rm II}=\int_{\R^{2N}}|\Psi(x_1,\dots,x_N)|^2 |\nabla f(x_i-x_j)| | \nabla f(x_k -x _j)|\prod_{l=1}^N dx_l
\eeq
for $i\neq j\neq k$. 

 It follows  from the definition \eqref{def:dell} that as a quadratic form on  $\B_\ell$
\beq\label{Dmap}
\D^{(\ell)}_{ij}= (\bar z_i-\bar z_j)^{-\ell} \D^{(0)}_{ij}(z_i-z_j)^{-\ell}.
\eeq
Moreover, in the notation of \cite{LS},
\beq \label{D0} 
\D^{(0)}_{ij}=\delta_{ij}.
\eeq
Because of \eqref{Dmap} and \eqref{D0} we can rely on previous results proved in \cite{LS} for the case $\ell=0$. In fact, \cite[Lemma~2]{LS} implies\footnote{Lemma~2 in \cite{LS} is stated in three dimension, but an inspection of its proof shows that it equally holds in the two-dimensional case as well.} 
 that for  $\Psi \in \mathcal{B}_0$ and radial $g\geq 0$ 
\beq\label{lem2a}
\int_{\R^{2N}} |\Psi(x_1,\dots,x_N)|^2\ g(x_i-x_j)  \prod_{l=1}^N dx_l \leq \int_{\R^2} g \, \langle \Psi | \delta_{ij} \Psi\rangle + C \int_{\R^2} g(x) \frac{|x|^4}{1+|x|^4} dx \, \| \Psi\|^2 
\eeq
for some constant $C>0$, as well as
\beq\label{lem2b}
\int_{\R^{2N}}|\Psi(x_1,\dots,x_N)|^2 g(x_i-x_j)  g(x_k -x _j) \prod_{l=1}^N dx_l \leq C \left( \int_{\R^2} g \right)^2 \|\Psi\|^2 
\eeq
for $i\neq j\neq k$. 

We define 
\beq g(x)=|x|^{2\ell}(|\nabla f(x)|^2+ \tfrac 12v_{a_n}(|x|)|f(x)|^2)\eeq
and write I as 
\beq\label{ft1}
\int_{\R^{2N}} |\Psi_{ij}(x_1,\dots,x_N)|^2g(x_i-x_j)\prod_{k=1}^N dx_k
\eeq
where $\Psi_{ij}$ stands for $\Psi$ with a factor $(z_i-z_j)^\ell$ canceled. Note that by \eqref{Dmap} we have
\beq\langle \Psi_{ij}|\D^{(0)}_{ij}\Psi_{ij}\rangle=\langle \Psi|\D^{(\ell)}_{ij}\Psi\rangle\eeq
and by \eqref{def:bell} 
\beq\label{2ndt}
\int_{\mathbb R^2} g =a_n^{2\ell}\, 4\pi\ell  b_\ell.\eeq
When applying \eqref{lem2a}, the first term gives after summation over $ij$ the desired bound
\beq\label{firstterm}
a_n^{2\ell}\, 8\pi\ell  b_\ell \,\langle \Psi| \hf_\ell\Psi\rangle.
\eeq
To bound the second term, 
we note that if $v$ is supported in $\{|x|\leq R_0\}$ then $v_{a_n}$ is supported in $\{|x|\leq a_n R_0\}$.  Pick an $R>R_0$ and split the integral in the last term in \eqref{lem2a} into an integral over $\{|x|<a_nR\}$ and a remainder where $v_{a_n}=0$. The first part of the integral is bounded by
$(\int g) R^4a_n^4$.   For $|x|>a_nR$ we have $f(x)=(1-a_n^{2\ell}b_\ell/|x|^{2\ell})$
and thus $|\nabla f(x)|\sim a_n^{2\ell}/|x|^{2\ell+1}$ so
$\int_{|x|>a_nR} |x|^{2\ell}|\nabla f(x)|^2 dx\sim a^{2\ell} R^{-2\ell}$.
With $R=a_n^{-2/(\ell+2)}$ we see that the integral is smaller than $\int g$ by a factor $a_n^{4\ell/(\ell+2)}$.
 The whole term I  therefore gives  \eqref{firstterm} as the leading contribution. 

A bound on II can be obtained with the aid of  \eqref{lem2b}. The resulting bound  is of higher order than $\int_{\mathbb R^2} g$ (in fact,  of the order $(\int_{\mathbb R^2} g)^2$)  and  vanishes upon multiplication by $a_n^{-2\ell}$ in the limit $a_n\to 0$.
Altogether we thus obtain
\beq
\limsup_{n\to \infty} a_n^{-2\ell} F_{a_n}(\Psi_n) \leq 4\pi \ell b_\ell \sum_{i\neq j} \langle \Psi | \D^{(\ell)}_{ij} \Psi\rangle = 8\pi \ell b_\ell \, G^{(\ell)}(\Psi).
\eeq
In combination with the lower bound, this concludes the proof of Theorem~\ref{thm:main} for $\ell \geq 1$. 

\subsection{The $\ell=0$ case}\label{ss3.5}
 The special feature of the $\ell=0$ case in two dimensions is that the minimizer $f_0$ of  \eqref{def:b0} depends on $R$,  and the minimal value of \eqref{def:b0} depends logarithmically on the parameters. By choosing $R$ appropriately, the basic proof strategy goes through also for $\ell =0$, with only minor modifications compared to the case $\ell \geq 1$. 
Effectively, one is lead to 
the replacement
\beq\label{subst} \ell b_\ell a^{2\ell}\rightarrow \frac 1{\ln(R^2/a^2 b_0^2)} =\ln(1/a^2)^{-1}(1+o(1)) 
\eeq
with $o(1)$ tending to zero as $a\to 0$ for fixed $R$. With this replacement the Dyson Lemma in Eq.~\eqref{eq:dyson} holds also for $\ell=0$, and the estimates for the lower bound are obtained in the same way as for $\ell\geq 1$, by first letting $a\to 0$ followed by $R\to 0$. 

For the upper bound the trial function $f$ is defined as $f_0(r/a_n)$ where $f_0$ minimizes \eqref{def:b0} with the choice $R= R'/a_n$ for some fixed $R'>0$ independent of $n$.  In particular $\nabla f(r)=0$ for $r\geq R'$. Again letting $R'\to 0$ after $a_n\to 0$ (in order for the last term in \eqref{lem2a} to be negligible compared to \eqref{2ndt}), we conclude the  
upper bound 
\beq\ln(1/a_n^2)^{-1}(1+o(1))\, 8\pi \,\langle \Psi| \hf_0\Psi\rangle\eeq in place of \eqref{firstterm}.

\section{Extension to three dimensions}\label{sec:3d}

In this section we shall show analogous results in the three-dimensional case, with a strong confinement in the third direction. Consider a potential $V:\R \to \R$ such that the Schr\"odinger operator $-\partial_u^2 + V(u)$ has a ground state $\chi \in H^1(\R)$ with corresponding energy $E$. 
On $L^2(\R^3)$, define $h\geq 0$ as 
\beq
h = ( -\rmi \nabla + x \wedge e_3)^2 + V(x^{(3)}) - 2 - E 
\eeq
where $e_3=(0,0,1)$ denotes the unit vector in the $x_3$-direction. For $v\geq 0$ radial and of compact support, we define 
\beq
H_a = \sum_{i=1}^N h_i + \sum_{i<j} v_a ( | x_i-x_j | )
\eeq
on $\Hh= L^2(\R^{3N})$, where $v_a(x) = a^{-2} v( x /a)$.

The relevant scattering parameters are now given by 
\begin{equation}\label{def:bell3}
b_\ell = \frac 1{4 \pi (2\ell+1) }  \min \left\{ \int_{\R^3} |x|^{2\ell} \left( |\nabla f(x)|^2 + \tfrac 12 v(x) |f(x)|^2 \right) dx \, : \, \lim_{|x|\to\infty} f(x) = 1\right\}
\end{equation}
for $\ell \in \N \cup \{ 0\}$. The corresponding minimizers $f_\ell$ satisfy $0\leq f_\ell \leq 1$, and  $f_\ell(x) = 1 - b_\ell/|x|^{2\ell+1}$ for  $|x|> R_0$. In particular, $b_\ell = R_0^{2\ell+1}$ for hard spheres.

We introduce  a sequence of closed subspaces
\beq
\Hh \supset \B_0 \supset \B_1 \supset \dots
\eeq
where $\B_\ell$ for $\ell \geq 0$  consists of $\psi \in \Hh$ of the form
\begin{equation}\label{potf}
\Psi(x_1,\dots,x_N) =  \varphi(z_1,\dots,z_N) \prod_{i<j} (z_i - z_j)^\ell  \prod_{k=1}^N \chi(x^{(3)}_k) \rme^{-\tfrac 12  |z_k|^2  } 
\end{equation}
with $ \varphi : \C^N\to \C^N$ analytic. We write $x =(x^{(1)}_j, x^{(2)}_j, x^{(3)}_j)$ and identify $(x^{(1)}_j, x^{(2)}_j) \in \R^2$ with  $z_j = x^{(1)}_j + \rmi x^{(2)}_j\in \C$. 
Note that again $\B_0$ coincides with  the kernel of $H^{(0)} = \sum_{i=1}^N h_i$, since for $\Psi(x) = \rme^{- |z|^2/2 } \chi(x^{(3)}) \phi(x)$, 
\begin{equation}\label{eq:rep3}
\langle \Psi | h \Psi \rangle = \int_{\R^3} \rme^{-|z|^2} |\chi(x^{(3)})|^2 \left(  4\left| \partial_{\bar z} \phi(x) \right|^2  + \left| \partial_{x^{(3)}} \phi(x) \right|^2 \right) dx.
\end{equation}
Obviously the spaces $\B_\ell$ can be naturally identified with the corresponding spaces in two dimensions, by simply multiplying functions in the latter spaces by $\prod_{k=1}^N \chi(x^{(3)}_k)$. In particular, the operators $\D^{(\ell)}$ naturally act on $\B_\ell$ in the same way as in the two-dimensional case, and similarly for $\hf_\ell$. 

Let $c_0 = 1$ and 
\beq
c_\ell = \frac{\sqrt{\pi}}2 \frac{\Gamma(1 + \ell)}{\Gamma(3/2 + \ell)}  = \prod_{j=1}^\ell \frac {2j}{2j +1}
\eeq
for $\ell \geq 1$. Let also $\tilde b_\ell = b_\ell c_\ell$. 

\begin{theorem}\label{thm:3d}
For any $\ell\geq 0$, $a^{-2\ell-1} H_a$ converges to $8\pi (2\ell + 1) \tilde b_\ell \int |\chi|^4 \, \hf_\ell$ as $a\to 0$ in the strong resolvent sense, i.e., for any $\mu > 0$ and $\Psi \in \Hh=L^2(\R^{3N})$ 
\beq
\lim_{a\to 0} \left( \mu + a^{-2\ell-1} H_a \right)^{-1} \Psi  =  \left(\mu + 8\pi (2\ell+1) \tilde b_\ell \mbox{$\int |\chi|^4$} \,  \hf_\ell  \right)^{-1} P_\ell \Psi
\eeq
strongly in $L^2(\R^{3N})$, where $P_\ell$ denotes the projection onto  $\B_\ell\subset \Hh$.  
\end{theorem}

The proof of Theorem~\ref{thm:3d} proceeds along very similar lines as the one in the two-dimensional case, hence we will not present it in full detail here. We merely point out the main differences. 

The relevant pre-factor $(2\ell+1) \tilde b_\ell$ naturally arises through
\begin{equation}\label{def:bell3n}
4 \pi (2\ell+1) \tilde b_\ell =  \min \left\{ \int_{\R^3} |z|^{2\ell} \left( |\nabla f(x)|^2 + \half  v(x) |f(x)|^2 \right) dx \, : \, \lim_{|x|\to\infty} f(x) = 1\right\}
\end{equation}
with a factor $|z|^{2\ell}$ in place of  $|x|^{2\ell}$ in \eqref{def:bell3}. 
To see the validity of \eqref{def:bell3n}, note that for radial functions $f$ we have
\beq
|z|^{-2\ell} \nabla |z|^{2\ell} \nabla f = |x|^{-2\ell} \nabla |x|^{2\ell} \nabla f
\eeq
hence the minimizer of \eqref{def:bell3} is also the minimizer of \eqref{def:bell3n}. The factor $c_\ell$ then results from averaging $|z|^{2\ell}$ over the unit sphere.  

As for Theorem~\ref{thm:main}, Theorem~\ref{thm:3d} is proved by showing $\Gamma$-convergence of the corresponding quadratic forms. 
The upper bound follows in the same way as in two dimensions, using \eqref{def:bell3n}. Note that by definition
\begin{equation}
\int_{\R^{3N}} \frac{ |\Psi(x_1,\dots,x_N)|^2}{|z_i-z_j|^{2\ell}} \delta(x_i-x_j) \prod_{k=1}^N dx_k = \int_\R |\chi|^4 \, \langle \Psi|\D^{(\ell)}_{ij}\Psi\rangle
\end{equation}
for $\Psi \in \B_\ell$, which explains the additional pre-factor $\int |\chi|^4$.

 For the lower bound, the relevant Dyson Lemma reads as follows:

\begin{lemma}
For any $\ell \geq 0$ and $R>a R_0$, and any $y = (\eta,y^{(3)}) \in \C \times \R \equiv \R^3$,  we have
\begin{align}\nonumber
&\int_{|x-y| < R}  \rme^{-|z|^2} | \chi(x^{(3)})|^2 \left( 4 |\partial_{\bar z} \phi(x)|^2 + |\partial_{x^{(3)}} \phi(x)|^2 + \tfrac 12 v_a(|x-y|) |\phi(x)|^2 \right) dx 
\\ \nonumber & \geq 4\pi (2 \ell +1)  \tilde b_\ell a^{2\ell+1}  \rme^{|\eta|^2 -R^2 } \min_{|x^{(3)}|<R} |\chi(y^{(3)}-x^{(3)})|^2   \\ & \quad  \times \left|   \frac 1{4\pi c_\ell R^{2(\ell +1)} } \int_{|x-y|=R}   \rme^{-z \bar \eta} (\bar z-\bar \eta)^{\ell} \phi(x) d\sigma \right|^2 \label{eq:dyson3}
\end{align}
where $\sigma$ denotes the surface measure on the sphere.
\end{lemma}

Its proof can be obtained by following the arguments in the two-dimensional case line by line.  Note that if $\varphi$ is independent of $x^{(3)}$ and analytic in $z$, we have
\beq
\frac 1{4\pi c_\ell R^{2(\ell +1)} } \int_{|x-y|=R}   e^{-z \bar \eta} (\bar z-\bar \eta)^{\ell} \phi(z) d\sigma =  \frac 1{2\pi \rmi} \oint_{|z-\eta|=R}  e^{-z \bar \eta}\frac{  \phi(z)}{(z-\eta)^{\ell +1}} dz
\eeq
where $dz$ is again the complex line element. The right side is independent of $R$ in this case, and coincides with the corresponding expression in two dimensions. Using that, as an $H^1$-function, $\chi$ is uniformly H\"older continuous, one easily sees that  
\beq
\lim_{R\to 0}  \int_\R |\chi(t)|^2  \min_{|x|<R} |\chi(t-x)|^2 dt = \int_\R |\chi|^4.
\eeq
 The remainder of the proof of the lower bound then proceeds as in Section~\ref{ss:lb}.

Assuming the potential $V$ to be confining, i.e., $\lim_{t\to \pm\infty} V(t) = + \infty$, and adding an additional confining potential $\lambda \sum_{i=1}^N |z_i|^2$ in the perpendicular directions, the analogue of Corollary~\ref{cor1} can be seen to hold in the three-dimensional case as well. The relevant Laughlin wavefunctions are simply given by \eqref{def:Lau} multiplied by $\prod_{k=1}^N \chi(x_k^{(3)})$, the ground state in the direction of the magnetic field.

\bigskip
{\it Acknowledgments.} The work of R.S. was supported by the European Research Council (ERC) under the European Union's Horizon 2020 research and innovation programme (grant agreement No 694227). J.Y. gratefully acknowledges hospitality at the LPMMC Grenoble and valuable discussions with Alessandro Olgiati and Nicolas Rougerie.

\end{document}